# V2V Spatiotemporal Interactive Pattern Recognition and Risk Analysis in Lane Changes

Yue Zhang[a], Yajie Zou[a,*], Lingtao Wu[b]

[a] key Laboratory of Road and Traffic Engineering of Ministry of Education, Tongji University, No. 4800 Cao'an Road, Shanghai 201804, China

[b] Texas A&M Transportation Institute, Texas A&M University System, 3135 TAMU College Station, Texas 77843-3135, USA

## ABSTRACT

In complex lane change (LC) scenarios, semantic interpretation and safety analysis of dynamic interactive pattern are necessary for autonomous vehicles to make appropriate decisions. This study proposes an unsupervised learning framework that combines primitive-based interactive pattern recognition methods and risk analysis methods. The Hidden Markov Model with the Gaussian mixture model (GMM-HMM) approach is developed to decompose the LC scenarios into primitives. Then the Dynamic Time Warping (DTW) distance based K-means clustering is applied to gather the primitives to 13 types of interactive patterns. Finally, this study considers two types of time-to-collision (TTC) involved in the LC process as indicators to analyze the risk of the interactive patterns and extract high-risk LC interactive patterns. The results obtained from The Highway Drone Dataset (highD) demonstrate that the identified LC interactive patterns contain interpretable semantic information. This study explores the spatiotemporal evolution law and risk formation mechanism of the LC interactive patterns and the findings are useful for comprehensively understanding the latent interactive patterns, improving the rationality and safety of autonomous vehicle's decision-making.

Key words: lane change, interactive pattern, traffic risk, driving primitive

## 1. Introduction

Lane change (LC) is a daily-routine but challenging driving task which includes vehicle movement in both longitudinal and lateral directions and close interaction between multiple vehicles. Every year, there are between 240,000 to 610,000 LC crashes reported to the police and more than 60,000 roadway users injured as a result of improper LC execution according to NHSTA(Fitch et al., 2009). With the advantages of perception and information acquisition, autonomous vehicles can significantly reduce traffic crashes. However, how autonomous vehicles make reasonable decisions in complex LC scenarios is a major challenge at present. On the one hand, human behavior is heterogeneous and stochastic. On the other hand, the behavior of each vehicle in the LC scenario is dynamically affected by surrounding vehicles. Therefore, it is necessary to understand the multi-vehicle interactive behavior patterns and reveal the interaction mechanism to facilitate autonomous vehicles to make reasonable decisions.

* Corresponding author.
E-mail addresses: zhangyue18@tongji.edu.cn (Yue Zhang); yajiezou@hotmail.com (Yajie Zou); wulingtao@gmail.com (Lingtao Wu).

As one of the most important microscopic behaviors of traffic flow, a large number of studies have been conducted to analyze LC behavior. For example, Woo et al. (2017) defined four LC phases according to the relationship between the vehicle and the center line and other features, namely keeping, changing, arrival and adjustment. Ni et al. (2020) divided LC into two stages. In the first stage, the subject vehicle completes LC operation involving longitudinal and lateral movements. The second stage only involves the adjustment of longitudinal headway of subject vehicle on the target lane. These studies assumed that the LC behavior is not affected by surrounding vehicles. In order to improve the accuracy of the LC behavior simulation, some studies took into account the impact of surrounding environment on the subject vehicle. Previously, Gipps (1986) considered adding safety gap and presence of heavy vehicles to the LC decision. Later, various factors indicating surrounding vehicles were considered, such as relative speed(Tang et al., 2018; Tang et al., 2019), gap acceptance(Hidas, 2005), and turning signal(Ponziani, 2012), etc. However, these studies only focused on the impact of other vehicles on the subject vehicle, but ignored the impact of the subject vehicle on surrounding vehicles. In order to consider the interaction of vehicles in LC scenarios, some studies classified LC scenarios based on prior knowledge. Hidas (2005) categorized LC into three classes: free, forced, and cooperative(Wang et al., 2019). Halati et al. (1997) classified LC maneuvers as mandatory lane changes, discretionary lane changes, and random lane changes. Due to the uncertainty of behaviors, the multi-vehicle interactive patterns are dynamically changing. In order to capture the impact of traffic context, Wang et al. (2020) proposed a real-time multi-vehicle collaborative learning approach to model spatial and temporal information among multiple vehicles. A few researchers introduced game theory to simulate the LC interactive behaviors(Ji and Levinson, 2020a; Ji and Levinson, 2020b). However, these real-time decision-making methods heavily reply on the sensor data, and it is difficult to explain the mechanism of interaction resulting in incomprehensible decisions. Therefore, it is necessary to explore the interactive patterns as a prior information for decision-making in advance.

Because the LC scenario has complex and high-dimensional features, it is challenging to extract interactive patterns from it. Some researchers addressed this issue by decomposing the entire scenarios into finite segments. For instance, the Gaussian mixture model (GMM) was applied to split the intersection encounter scenarios into discrete segments (Gadepally et al., 2014; Havlak and Campbell, 2014). The stochastic processes such as the Markov model (Tang et al., 2016; Zhang and Wang, 2019) can be used to describe them. This solution are conducive to improve learning performance from massive data. Zhang et al. (2021) proposed a primitive-based framework to learn interactive patterns by segmenting the LC scenarios. However, the study did not analyze the risk of interactive patterns, which may lead to autonomous vehicles learning improper operation and making dangerous decisions.

In recent years, surrogate safety measures have been widely used to quantify the risk of micro-behavior and severity of interaction. Saunier et al. (2010a) used some surrogate safety indicators for the analysis of vehicle-to-vehicle (V2V) spatial interactions at intersections, and interactions are classified into four categories: head-on, rear-end, side and parallel. Common surrogate safety indicators included Time-to-collision (TTC), Post-Encroachment Time (PET), Gap Time (GT), and Deceleration-to-Safety Time (DST)(Ni et al., 2016). Among these indicators, TTC is probably the most well-known and commonly applied for risk analysis (Jeong and Oh, 2017; Li et al., 2020; Nadimi et al., 2020; Noh and An, 2017). TTC refers to the time for a vehicle to collide with the preceding one without changing its current direction and speed. Wu et al. (2020) recognized real-time LC risk level based on TTC. According to the results of Laureshyn et al. (2010), the risk in an interactive scenario was not constant. For example, for a dangerous interactive scenario, some time periods may have high risk level, while some time periods may not have risk. The decomposition method also helps to focus on the most dangerous segment (interactive pattern) in a LC interactive scenario.

According to the previous studies, it is necessary to investigate the V2V spatiotemporal interactive patterns existing in the real LC scenarios based on the semantic decomposition method. Also, the risk level of these interactive patterns is conducive to enhance the safety of driving decisions. Motivated by these two issues, this study proposed a risk analysis framework for V2V interactive behavior patterns based on real-world trajectory data. This framework includes the interactive pattern recognition, semantic interpretation and risk analysis. The contributions of this study are as follows:

1. Present an unsupervised framework that integrates primitive-based interactive pattern recognition method and risk analysis method.

2. Establish the mapping relationship between the LC interactive scenarios and the real world, and analyze the spatiotemporal evolution mechanism of the interactive patterns such that autonomous vehicles can better understand human behavior.

3. Consider different interaction types between two vehicles to calculate TTC and identify high-risk LC interactive patterns.

The rest of this paper is organized as follows: Section 2 introduces the methodology used in this study. Section 3 documents the data collection and preprocessing. Section 4 discusses the learning process of models and analyzes the results of proposed method. Section 5 provides the conclusion and future work of this study.

## 2. Methodology

The framework proposed in this paper is shown in Figure 1. Firstly, specific LC scenarios are extracted based on naturalistic driving data. Then the primitives are segmented and clustered to obtain interactive patterns. Finally, semantic explanation and risk analysis of the interactive pattern are conducted.

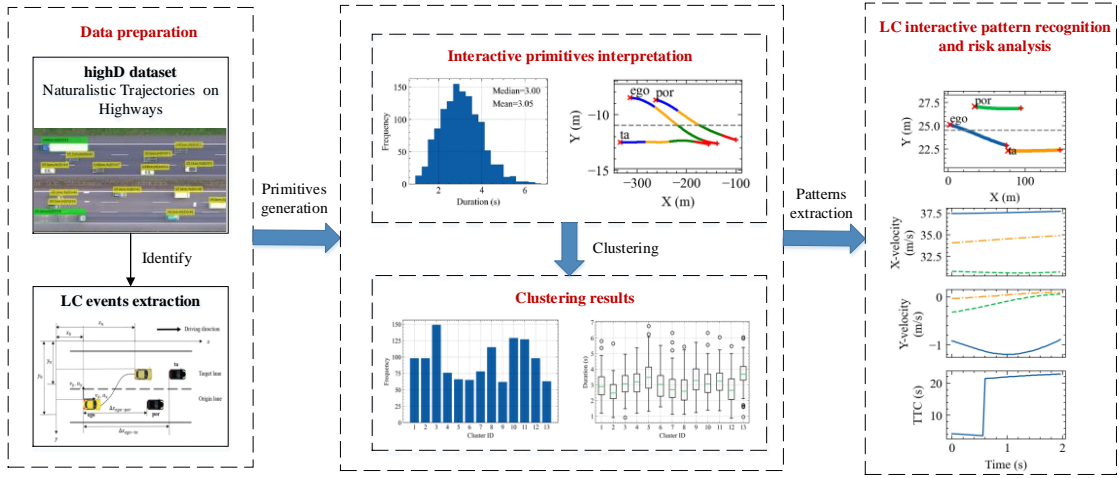

Figure 1  Framework of LC interactive patterns recognition and risk analysis

## 2.1 LC scenario and its primitives

The sequential LC event data includes the trajectories of three vehicles, expressed as

$$P = \{p_1, p_2, \cdots p_t, \cdots p_T\} \quad (1)$$

$P$ can be called a LC interactive scenario, where $p_t = \{x_t^{(1)}, y_t^{(1)}, x_t^{(2)}, y_t^{(2)}, x_t^{(3)}, y_t^{(3)}\} \in \mathbb{R}^6$. $x_t^{(1)}$, $x_t^{(2)}$ and $x_t^{(3)}$ are the longitudinal position of three vehicles at time $t$, respectively. $y_t^{(1)}$, $y_t^{(2)}$ and $y_t^{(3)}$ are the lateral position of three vehicles at time $t$. $T$ represents the time length of a LC scenario $S$. The primitive of $S$ is formulated as Eq.(2).

$$P_i = \{p_m, \cdots p_n\} \quad (1 \leq m \leq n \leq T) \quad (2)$$

Where $P_i \subseteq P$, $i$ is the $i$-th primitive.

## 2.2 Segment LC interactive primitives

LC interactions are dynamic process with stochastic human behavior. To analyze the complex scenarios with high-dimensional data, the GMM-HMM is introduced to decompose scenarios into semantic primitives. The GMM-HMM has been widely used in natural language processing (NLP) and automatic speech recognition (ASR) to process sequence data and mine the hidden state such as phonemes. The HMM in the GMM-HMM has a strong advantage in modeling dynamic behavior. In order to establish the relationship between driving scenarios and types of interactive patterns, the GMM is used to model the state-output distribution on the basis of HMM. The mixture component of GMM corresponds to the hidden state in the HMM. The entire process of the model is illustrated in Figure 2. The components of the GMM-HMM are described in detail below.

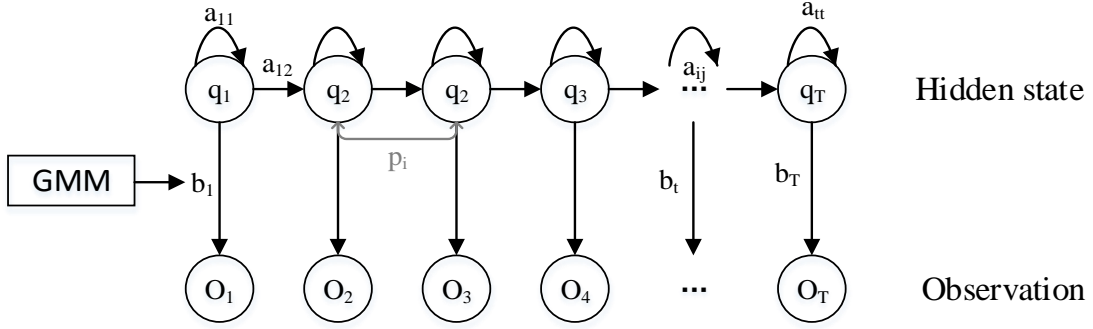

Figure 2 The schematic diagram of the proposed GMM-HMM model.

2.2.1 Gaussian Mixture Model

The trajectory data in LC interactive scenario are continuous, and the observation is heterogeneous due to the existence of multiple interactive patterns. Due to the strong ability of fitting multiple distributions and mining heterogeneity in data, the GMM is selected to model the emission probability in the HMM so as to establish the relationship between interactive patterns and LC interactive scenarios. The joint probability density function generated by multi-dimensional GMM is

$$P(p_t) = \sum_{i=1}^{N} \omega_i N(p_t | \mu_i, \Sigma_i)$$
$$= \sum_{i=1}^{N} \omega_i \frac{1}{(2\pi)^{d/2} |\Sigma_i|^{1/2}} \times \exp\left\{-\frac{1}{2}(p_t - \mu_i)^T (\Sigma_i)^{-1} (p_t - \mu_i)\right\} \quad (3)$$

$$\sum_{i=1}^{N} \omega_i = 1 \quad (4)$$

Where $N(p_t | \mu_i, \Sigma_i)$ is the $i$-th multivariate Gaussian distribution (dimension $d=6$) and $p_t = \{x_t^{(1)}, y_t^{(1)}, x_t^{(2)}, y_t^{(2)}, x_t^{(3)}, y_t^{(3)}\}$ is the observation of the LC interactive scenario at time $t$. $N$ is the number of components. $\mu_i$ is mean of the $i$-th Gaussian component, and $\Sigma_i$ is covariance matrix. $\omega_i$ is weight of the $i$-th component. In Eq.(3), $\mu_i$ and $\Sigma_i$ are the parameters that needs to be estimated in the GMM. The Expectation maximization (EM) algorithm is used for estimating these parameters. The convergence criterion is that the difference between the log likelihood values of two adjacent iteration steps is less than $10^{-10}$.

## 2.2.2 HMM and decoding

In the GMM-HMM, each component of the GMM is regarded as the hidden state of the HMM. The aim is to segment primitives of a given observation sequence $O$ using HMM. Some concepts and symbols involved in the HMM should be defined in advance.

- Hidden states: we represent the hidden states as $S = \{S_1, S_2, \cdots, S_N\}$, where $N$ is the number of hidden states and $q_t \in S$ is the state at time $t$. Although the states are hidden, there are often physical meaning in applications(Rabiner, 1989). In this study, the states are described as types of interactive patterns.

- Observations: the observable state sequence is $O = \{o_1, o_2, \cdots, o_t, \cdots, o_T\}$, $O_t = p_t = \{x_t^{(1)}, y_t^{(1)}, x_t^{(2)}, y_t^{(2)}, x_t^{(3)}, y_t^{(3)}\}$ is the observation at time $t$.

- Initial state distribution: $\pi = \{\pi_i\}$, where $\pi_i = P[q_1 = S_i]$, $1 \leq i \leq N$. $\pi_i$ is the probability which the Markov chain will start in state $i$.

- Emission probabilities: $B = \{b_j(o_t)\}$ are the sequence of probability of an observation $o_t$ generated from state $j$. In the GMM-HMM, the emission probability is generated by Eq.(3).

- State transition probability distribution: $A = \{a_{ij}\}$ is the state transfer matrix, which represents the probability of transition from state $i$ to state $j$. And the transfer probability for ($i,j$) pairs can be expressed as Eq.(5).

$$a_{ij} = P(q_{t+1} = S_j | q_t = S_i), \text{ s.t. } i \geq 1, j \leq N \text{ and } \sum_{j=1}^{N} a_{ij} = 1 \ \forall i \tag{5}$$

Thus, the HMM can be expressed $\lambda = (N, T, A, B, \pi)$, where $N$ is a hyper-parameter. As one of the three classical HMM problems, decoding problem is the core of composing LC interactive primitives. The decoding problem is how to find the optimal hidden state sequence given the observation sequence $O$ and model $\lambda$, which can be formulated as Eq.(6).

$$\gamma_t(i) = P(q_t = S_i | O, \lambda) \tag{6}$$

Based on the forward-backward algorithm, Eq.(6) can be transformed into Eq.(7).

$$\gamma_t(i) = \frac{\alpha_t(i)\beta_t(i)}{P(O|\lambda)} = \frac{\alpha_t(i)\beta_t(i)}{\sum_{i=1}^{N} \alpha_t(i)\beta_t(i)} \tag{7}$$

Where $\alpha_t(i)$ accounts for a part of observations $\{o_1, o_2, \cdots o_t\}$ and $\beta_t(i)$

accounts for the residual observations $\{o_{t+1}, o_{t+2}, \cdots o_T\}$ given $S_i$ at time $t$. $P(O|\lambda)$ is the normalization factor to make sure

$$\sum_{i=1}^{N} \gamma_t(i) = 1 \tag{8}$$

Then most likely state $q_t$ can be solved as Eq.(9).

$$q_t = \underset{1 \leq i \leq N}{\arg\max} [\gamma_t(i)], \ 1 \leq t \leq T \tag{9}$$

According to the most likely state sequence $Q = \{q_1, q_2 \cdots q_t\}$ obtained by Eq.(9), if the state sequence corresponding to the observation value is continuous and consistent, these observations are segmented into a primitive (because they contain the same semantic information). The Viterbi algorithm(Rabiner, 1989) is used to find the best hidden state sequence in the GMM-HMM. Viterbi is a dynamic programming method, which is used to search the most probable path by taking the maximum probability value of all possible previous hidden state sequences.

**2.3 Clustering of interactive patterns**

For the large number of primitives extracted by the GMM-HMM, this study uses a clustering method to separate them into homogeneous interactive patterns.

2.3.1 Data scaling and normalization

The primitives extracted from the GMM-HMM has considerably different length. Before applying the clustering model, these primitives should be scaled into the same length $l$. This study uses linear interpolation to scaling-down and scaling-up, which ensures that the scaled trajectory is similar to the original trajectory. Given the data point $p_0$ at time $t_0$ and $p_1$ at time $t_1$, the unknown data point at time $t \in (t_0, t_1)$ can be calculated by Eq.(10).

$$p_t = p_0 + (t - t_0) \frac{p_1 - p_0}{t_1 - t_0} \tag{10}$$

Since the dimensions of longitude and latitude of the vehicle trajectory are different, normalization is a necessary step. Therefore, each sample of the input sequences is standardized using Min-Max normalization which brings $p_t$ to within a standard [-1, 1] range.

2.3.2 DTW distance based K-means clustering

The K-means clustering can lead excellent results in terms of unsupervised feature

learning(Coates and Ng, 2012). The K-means is used to cluster LC interactive patterns. The primitive is a time series composed of the temporal feature of the vehicle trajectory. It is meaningless to simply calculate the center of a trajectory using the traditional K-means clustering method. Therefore, this study proposes to use the K-means clustering method based on Dynamic Time Warping (DTW) to classify the categories according to the similarity between time series.

DTW is a shape-based similarity measure for sequence data. DTW distance is a length of the optimal alignment between two given primitives which are time-series data. It uses dynamic programming to find an optimal path with a minimum distance between two given time series.

Before clustering, the input primitive $P_i \in \mathbb{R}^{6 \times l}$ should be reshaped as $P_i \in \mathbb{R}^{1 \times 6l}$.

$$P_i = [p_1, p_2 \cdots p_a, \cdots p_l], \ p_a \in \mathbb{R}^6 \tag{11}$$

Then the DTW distance based K-means method is used to cluster primitives. Assuming all primitives $P = \{P_1, P_2, \cdots P_N\}$ are grouped into $N$ clusters $C = \{C_1, C_2, \cdots C_K\}$ and the center of each cluster is $\mu_i$. The goal of the proposed method is to minimize the within-cluster sum-of-squares by Eq.(12).

$$\lambda_w = \min \sum_{i=1}^{K} \sum_{P \in C_i} \|P_i - \mu_i\|^2 \tag{12}$$

Finally, the primitives with similar temporal and spatial features are grouped into a cluster.

**2.4 Interactive pattern risk calculation**

In order to assess the risk of each interactive pattern and identify high-risk interactive patterns, a risk calculation method based on TTC is introduced in this section. The TTC at time *t* is defined as the remaining time between two vehicles driving in the current state until a collision. The main reasons that TTC is selected as the risk indicator are: (1) it can intuitively reflect the driver's perceived risk, (2) it exists in all types of interactions, and (3) it evolves continuously during each interaction (Saunier et al., 2010b; St-Aubin et al., 2013).

In LC scenarios, there are two types of interaction between two vehicles according to the relative position and their lanes (Figure 3). The type A interaction refers to two vehicles driving in the same lane and the speed of the following vehicle is greater than that of the preceding vehicle. The type C interaction is vehicles in different lanes experiencing side interaction when their expected path cross and projected positions overlap. In type C, it is necessary to consider the situation that the vehicles reach the collision point in both the longitudinal and lateral directions at the same time. The method of calculating two types of TTC is based on the work conducted by (Laureshyn et al., 2010; St-Aubin et al., 2013) as well as some basic geometry calculations.

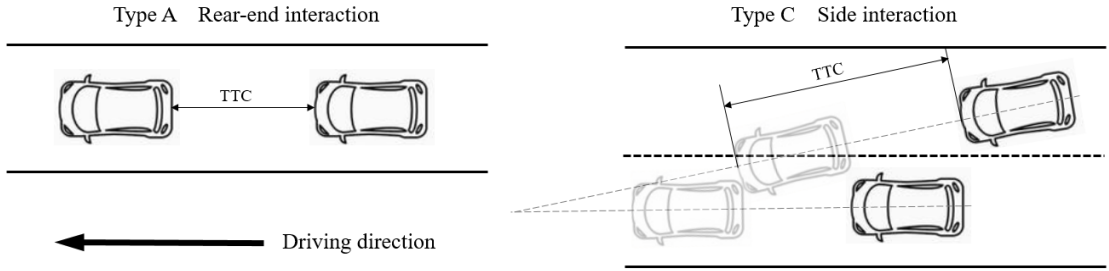

Figure 3 Interaction types between two vehicles in the LC interactive scenario.

After obtaining the TTC of the interaction between two vehicles at each moment, it is necessary to aggregate them as the primitive's risk which needs to cover the risk of all interaction pairs at all moments in the primitive. The risk of every two pairs is represented by a minimum TTC at each moment (one measure per interaction) and the risk of a primitive is represented by the mean TTC of three interactions (one measure per primitive).

## 3. Data description

### 3.1 Data collection

The data employed in this study are naturalistic driving trajectory dataset extracted from drone video called highD. It is a large-scale public dataset widely used in traffic flow modelling and safety analysis in recent years (Kruber et al., 2019; Kurtc, 2020; Pierson et al., 2019; Schneider et al., 2020; Wirthmüller et al., 2020). Compared with other naturalistic driving datasets, it has the characteristics of high accuracy. The position error does not exceed 10cm. The highD dataset was collected from six locations at German freeways around Cologne during 2017 and 2018 (Krajewski et al., 2018), as marked in Figure 4. The six data collection site include basic segments and merge segments. Due to differences in the geographic characteristics, the LC interactive behavior on the merge segments is quite different from the LC interactive behavior on the basic segments. Unfortunately, only three recordings out of all 60 records are in the merge segments. Considering the sufficiency and diversity of the data, five locations of basic segments are finally selected. Table 1 shows the information of all sites selected in this study. The data are collected at the frequency of 25 Hz.

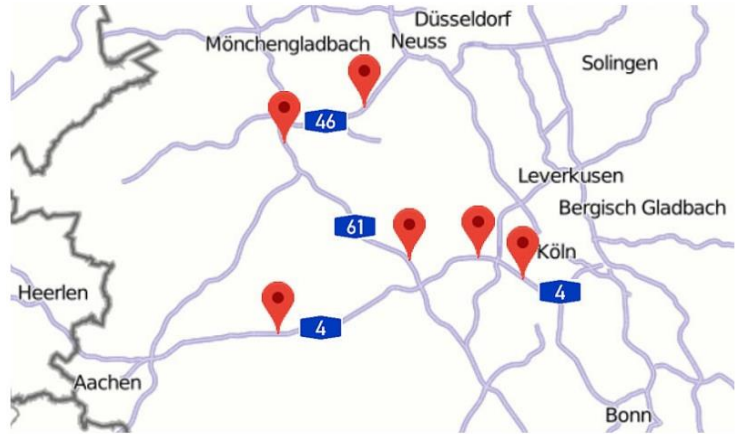

Figure 4　Locations of recordings included in highD (Krajewski et al., 2018)

Table 1 Description of the selected segments

| Location ID | Lane width (m) | Segment length (m) | Number of lanes (both direction) |
|---|---|---|---|
| 1 | 4.07; 3.88; 4.15; 4.24; 3.80; 4.15 | 420 | 6 |
| 2 | 4.08; 3.84; 3.96; 3.84 | 420 | 4 |
| 3 | 3.82; 3.65; 3.73; 3.74; 3.56; 3.91 | 420 | 6 |
| 4 | 3.97; 3.63; 3.62; 3.53; 3.80; 3.89 | 420 | 6 |
| 5 | 4.09; 3.84; 3.96; 3.97; | 420 | 4 |

1 The lane width column shows the width of each lane which separated by ";".
2 "-" in the speed limit column means there is no speed limit. The speed limit is the same for every driving lane.

### 3.2 Vehicle type classification

　　Due to different vehicle dynamic performance, the interactive patterns of different vehicle types should also be discussed separately. First of all, the vehicle types need to be classified. In order to discretize the continuous vehicle size data, this study applies the K-means clustering method to classify three vehicle types according to vehicle length and width, namely passenger cars (PC), heavy-duty vehicles (HV), and over-sized truck (OT). Figure 5 shows the distribution of three vehicle types, in which the red triangle symbol is the centroid of clustering.

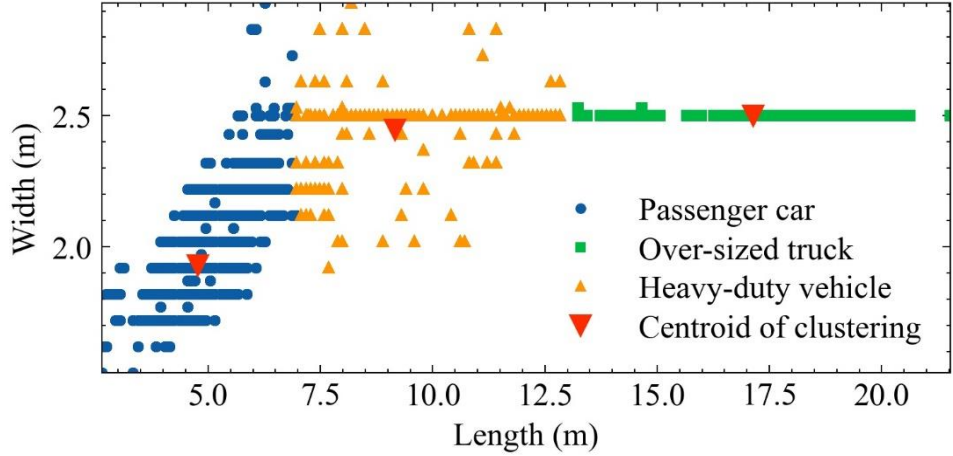

Figure 5 Clustering results of three vehicle types

**3.3 LC events extraction**

Figure 6 shows a common LC process. The lane where the LC vehicle starts to LC maneuver is called original lane, and the lane where the LC process is completed is called target lane. In order to explore the interaction between LC vehicle and the vehicle in the original lane and vehicle in the target lane, this study fixed the interested vehicles as three vehicles within a certain distance of LC vehicles. Note that the interactive patterns are not comparable when the number of vehicles in the place of interest is different. These three vehicles are the LC vehicle (*ego*), the vehicle in the original lane and the vehicle in the target lane (*ta*). Previous studies demonstrate that one of the most common types of LC scenario in which a vehicle changes lanes to pass a slower lead vehicle to maintain current speed or gain speed advantage (Bogard, 1999; Hetrick, 1997). Thus, the preceding vehicle in the original lane (*por*) is determined. The LC events can be automatically extracted by the following steps.

Step 1: Determine the *ego* vehicle. If the vehicle's driving lane ID changes (cross-lane), then the vehicle is marked as *ego* at this timestamp ($t_c$).

Step 2: Extract the complete LC process of the *ego* vehicle. According to the cross-lane timestamp recorded in step 1, the beginning and ending of LC process are searched forward and backward respectively in the trajectory record. We define the ending of LC as follows:

- Discontinuous increase or decrease of lateral position.

- $y_n - y_{t_c} > 0.9m$

- $a_{y_n} = a_{y_{n-1}}$

In the process of LC, trajectories before and after the cross-lane maneuver can be regarded as a symmetrical process. Therefore, the definition of the beginning and ending of LC are the same.

Step 3: Filter three-vehicle interactive scenario. The filter rules are $\Delta x_{ego-por} < 120m$ and $\Delta x_{ego-ta} \epsilon [-100m, 100m]$ at the beginning of LC.

Step 4: Classify interactive scenarios according to vehicle types. Based on the result of vehicle type classification, there are a total of 27 interactions between different vehicle types. In all interaction events, there are 578 *ego*(PC)-*por*(PC)-*ta*(PC) interaction events, 153 *ego*(PC)-*por*(OT)-*ta*(PC) interaction events, and 104 *ego*(PC)-*por*(HV)-*ta*(PC) interaction events. The frequency of other interactive types is less than 30. Therefore, the *ego*(PC)-*por*(PC)-*ta*(PC) interaction LC events are selected as the final samples in this study.

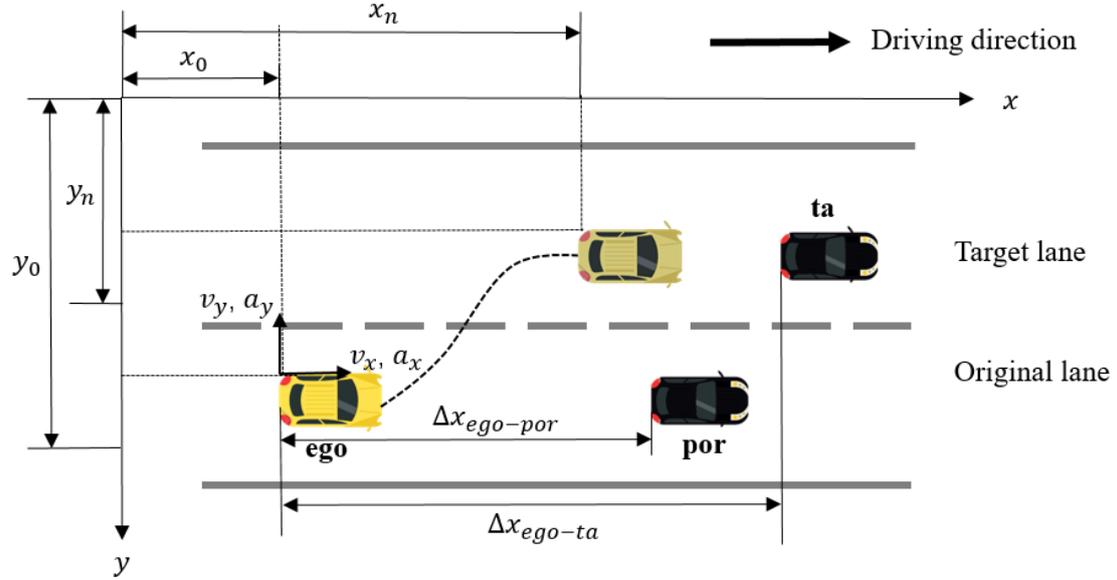

Figure 6 A schematic diagram of a typical LC interaction scenario

## 4. Results and analysis

### 4.1 LC interactive primitive extraction result

Firstly, the GMM-HMM is implemented to the 578 LC interactive events and generates 1224 primitives. In the training process of the GMM-HMM, the log-likelihood value is used to evaluate the learning performance of the model. In order to determine the optimal value of $N$ in Eq.(3) and Eq.(7), the value of $N$ is increased by 1 from 1 in the training process until the state sequence cannot be calculated. Because when $N$ increases, some state transition probabilities may be 0 so that there is no optimal state sequence. Among these $N$ values, the $N$ with the smallest log likelihood value of the model is selected. Figure 7 shows an example of the GMM-HMM learning process for one LC interactive scenario, which illustrates that the model has been trained to converge. Figure 8 shows distribution of the duration of all primitives. Short duration contains limited information, which cannot reflect the evolution of interactive pattern. Thus, the primitives with less than 10 frames (0.4s) are eliminated.

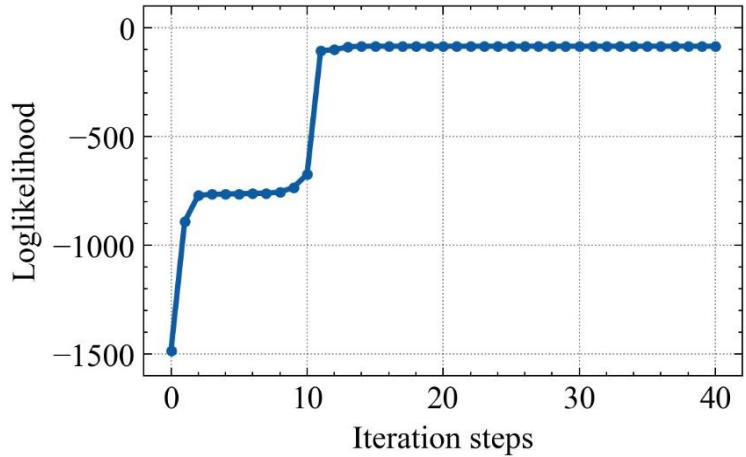

Figure 7 The learning process of the GMM-HMM to segment one of the LC interactive scenarios

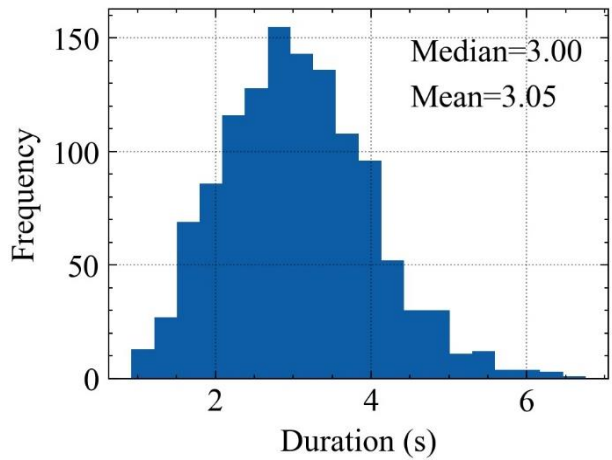

Figure 8 Distribution of LC interactive primitive duration

Figure 9 displays three common LC interactive events with extracted primitives. In order to clearly interpret the generated driving primitives, not only the trajectory but also the longitudinal speed, lateral speed and risk value are shown in top-down order in Figure 9. In the trajectory subgraph, "×" represents the starting of the trajectory, and "+" represents the ending of the trajectory. In the speed and risk value subgraphs, the solid line represents *ego* vehicles, the dashed line represents *por* vehicles, and the dotted line represents *ta* vehicles. Different colors represent different primitives in an LC interactive scenario, and the primitives in different LC interactive scenarios are independent.

The results show that the GMM-HMM can automatically divide segments containing different semantic information and identify the boundaries of primitives from complex scenarios. In detail, the results are summarized as four findings:
- The number and attribute of primitives in different LC interactive scenarios are significantly contrasting. For example, the LC interactive scenario of Figure 9 (b) consists of four primitives while that of Figure 9 (a) and (c) include two primitives. Note that the same color does not represent the same semantic segment in different events. For example, the red primitives in Figure 9 (b) and (c) are obviously different.
- The primitives are interpretable. Taking Figure 9(b) as an example, the *ego* and *por* vehicles change lanes at the same time, and the *ta* vehicle with higher longitudinal speed in the target lane is behind the *ego* vehicle at the beginning of LC. After LC process, all vehicles are located in front of the *ta* vehicle in the target lane. The entire interactive process is segmented into four primitives. In the blue primitive, *ego* and *por* vehicle start LC with increasing speed in lateral direction. The *por* vehicle has the lowest speed and the *ta* vehicle has the highest speed in longitudinal direction, and the *ta* vehicle starts to decelerate in order to avoid collisions. In the orange primitive, *ego* and *por* vehicles are approaching the lane line while adjusting the lateral speed. In the green primitive, *ego* and *por* vehicles cross the lane line to reach the target lane and decelerate in the lateral direction. At the same time, *ego* and *ta* vehicles quickly decelerate in the longitudinal direction in order to avoid collision. In the red primitive, *ego* and *por* vehicles driving steadily on the target lane and the lateral speed of them tends to zero. Due to the small gap between the *ego* vehicle and the *por* vehicle, the *ta* vehicle continues to decelerate until its speed is less than the speed of *ego* vehicle in longitudinal direction.
- The extracted primitives accurately distinguish the LC stage. In Figure 9 (b) and (c), the primitives are segmented before and after cross-line. When two vehicles are in LC process but not in the same stage (Figure 9 (a)), the primitives can also automatically make a trade-off between two vehicles' LC stage.
- The extracted primitives also have significant differences in risk level. In the process of V2V interaction, the risk level changes dynamically. The primitives can help us identify high-risk segments and the TTC value used to reflect their risk level. In the TTC graph of Figure 9 (b), the blue primitive show high risk, the orange primitive shows medium risk, the green primitive shows low risk, and red primitive shows no risk.

Therefore, the complex LC interactive scenarios can be decomposed into explanatory primitives automatically extracted by the GMM-HMM to facilitate a better understanding of the interactive patterns.

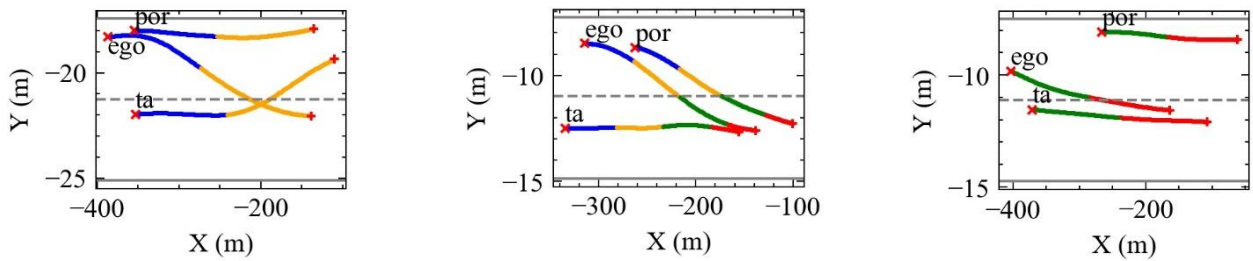

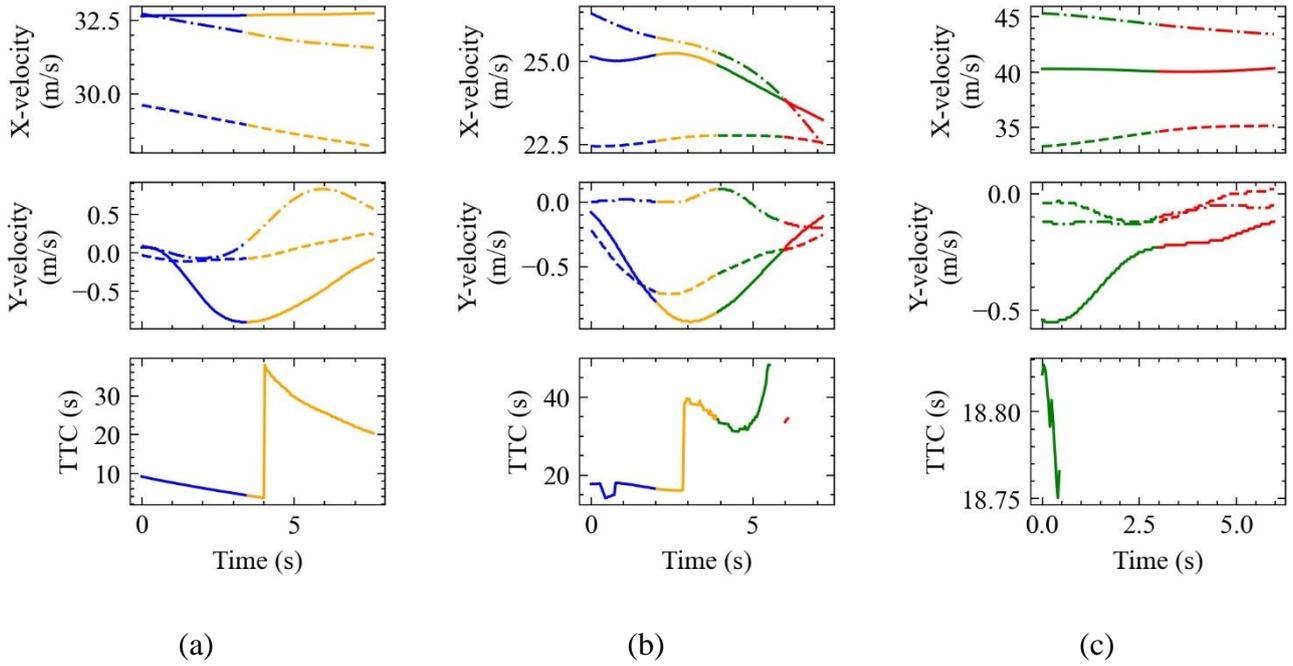

(a) (b) (c)

Figure 9 Decomposition results of three typical LC interactive scenarios.

## 4.2 Clustering evaluation

In order to summarize numerous primitives into finite interactive patterns, the DTW distance based K-means clustering is adopted to cluster the primitives. Before clustering, these primitives need to be scaled into the same length as the input of the clustering model. It is challenging to determine the optimal length of the input: a short length result in loss of the semantic information, while a long length increases the computational cost. In order to balance the learning performance against computational cost, this study chooses the median of all primitive durations (3s) as the expected primitive duration, as shown in Figure 8. So the input length in each primitive is scaled into 75. Figure 10 shows scaled data fit the original data very well, which can minimize the information loss of scaling-down.

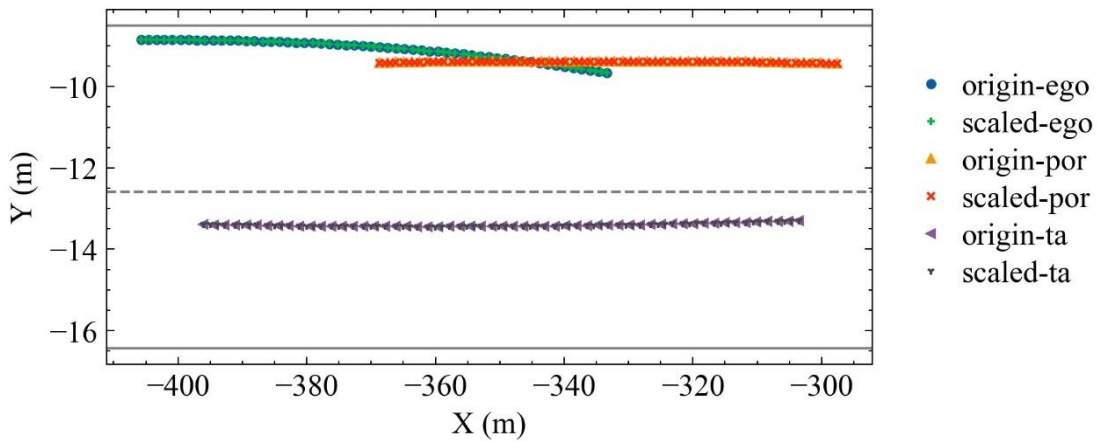

Figure 10 The scaled data and original data of one primitive

The number of clusters in the DTW distance based K-means clustering is determined based on the within-cluster sum-of-squares criterion. The within-cluster sum-of-squares criterion ($\lambda_w$) can be recognized as a measure of how internally coherent clusters are. The smaller the value is, the better the clustering result is. Figure 11 shows the results of $\lambda_w$, change rate of $\lambda_w$, and smoothed change rate of $\lambda_w$ after quadratic polynomial fitting. Considering the performance of the model and computational cost, *k*=13 is selected as the best number of clusters.

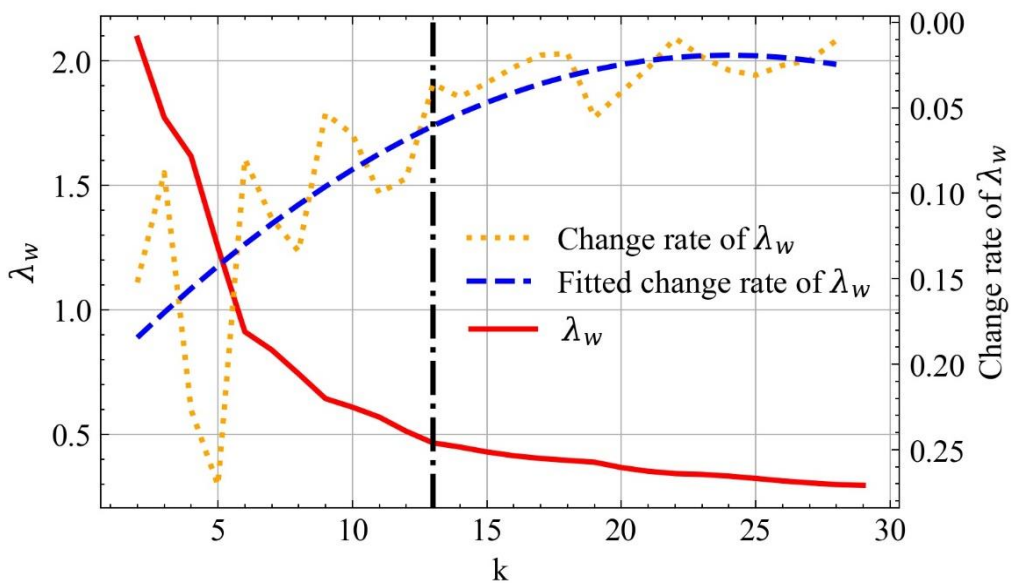

Figure 11 The curve of $\lambda_w$ over the number of clusters *k*

After clustering the 1,224 primitives, the frequency and duration distribution of each cluster of interactive patterns are shown in Figure 12 and Figure 13 respectively. In Figure 12, it is not difficult to find that cluster #3 is the most common interactive patterns. Cluster #9 and cluster #13 are the least common interactive patterns. The following sections focus on these three clusters of interactive patterns.

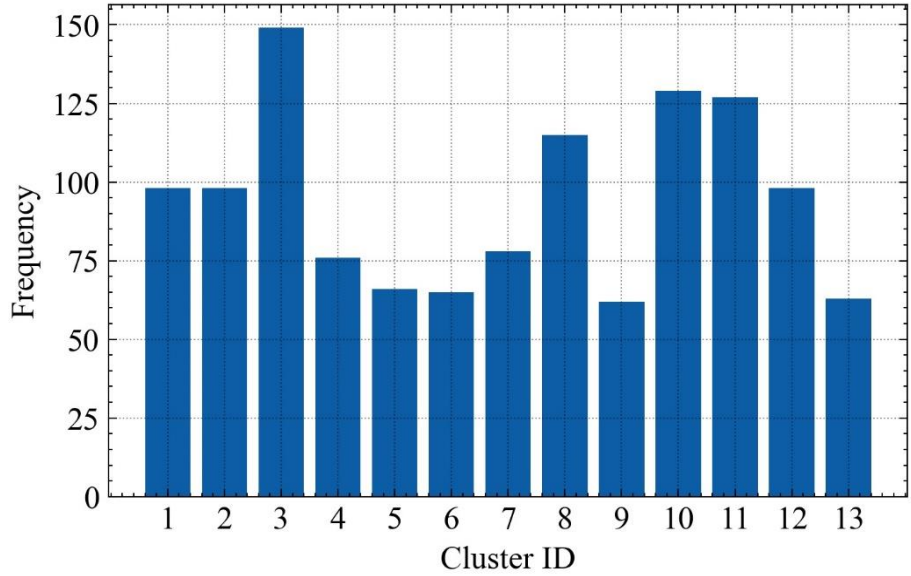

Figure 12 The frequency distribution of each cluster of interactive patterns

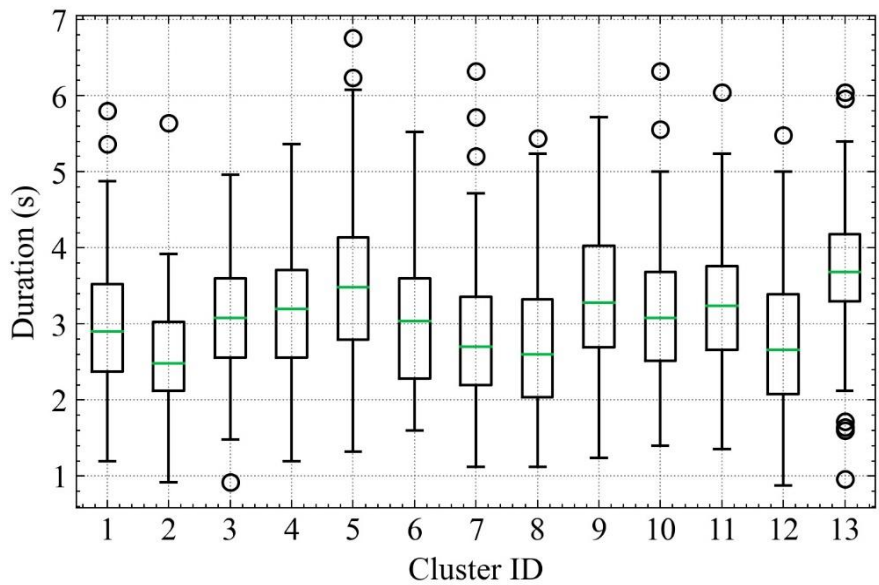

Figure 13 The duration distribution of each cluster of interactive patterns

**4.3 Interactive patterns analysis**

4.3.1 LC interactive pattern recognition and interpretation

Figure 14 (a) displays the most common interactive pattern (cluster #3). In the lateral direction, the *ego* vehicle just crossed the lane line and reduce its lateral speed with a constant acceleration, while the lateral speed of the *por* and *ta* vehicles keep around zero. In the longitudinal direction, the *ta* vehicle has the highest speed while the *por* vehicle has the lowest speed.

Figure 14 (b) and (c) represent the least common interactive pattern (cluster #9 and cluster #13).

Figure 14 (b) shows the *ta* and *ego* vehicles change lanes to each other's lanes respectively (their lateral speed directions are opposite), and they are both at the stage of crossing the lane line.

Figure 14 (c) indicates the adjustment stage after the completion of LC. In this interactive pattern, the lateral speed of the *ego* vehicle gradually decreases until it becomes stable, while that of the *ta* and *por* vehicles always keeps around 0. Due to the long adjustment stage in the target lane, cluster #13 rarely occurs in real traffic condition.

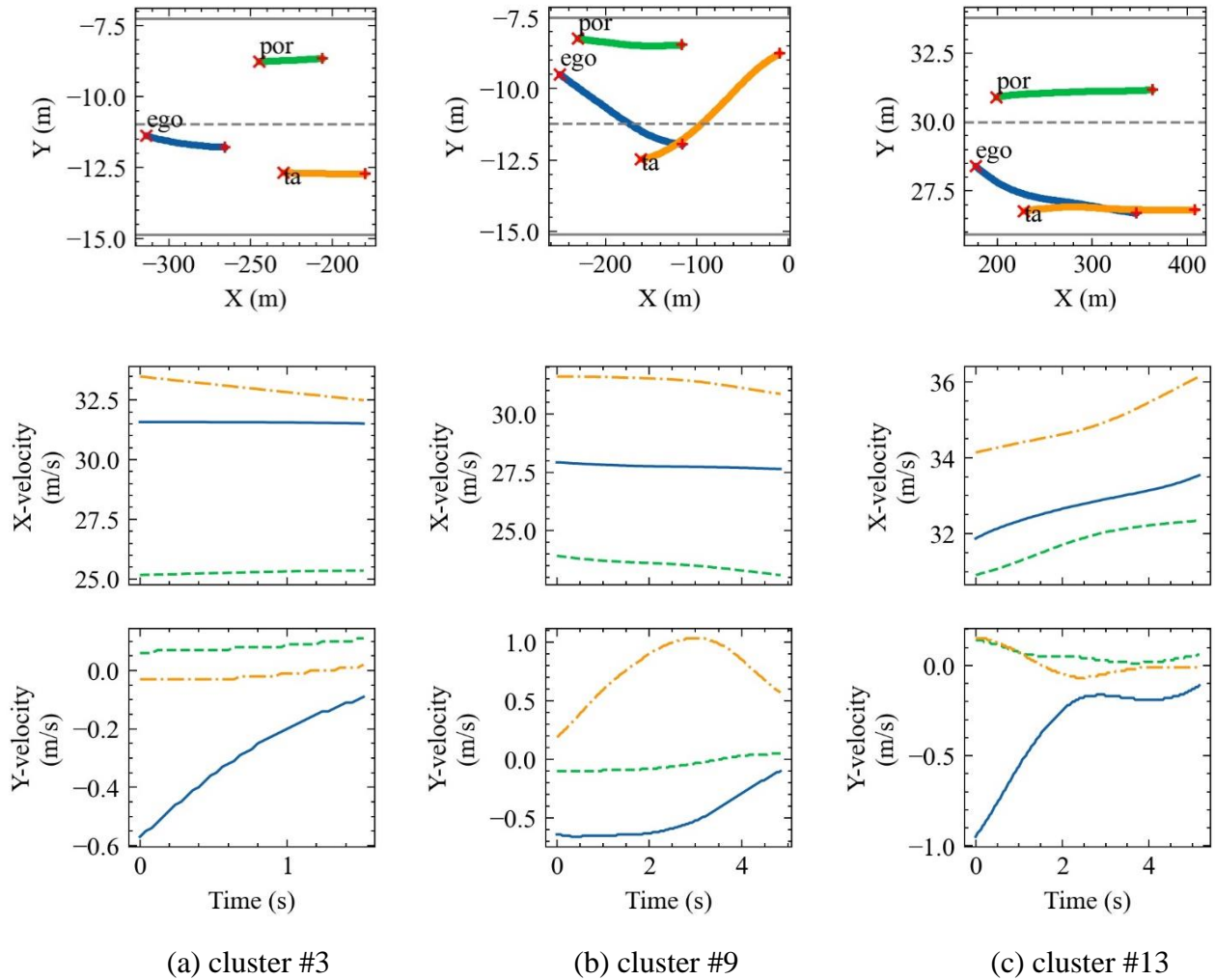

(a) cluster #3　　　　　　　(b) cluster #9　　　　　　　(c) cluster #13

Figure 14 Trajectory and velocity graphs of three clusters of primitives. In the trajectory subgraph, "×" represents the beginning of the trajectory, and "+" represents the ending of the trajectory. In the velocity subgraphs, the solid line represents *ego* vehicles, the dashed line represents *por* vehicles, and the dotted line represents *ta* vehicles.

4.3.2 LC interactive pattern risk analysis

According to the previous results, it was found that different primitives imply different risk levels. Further, the risk of interactive patterns are explored. The TTC value is selected to indicate both the risk of interactive patterns and the degree of interaction(Wirthmüller et al., 2020). Figure 15 shows the distribution of each cluster of interactive patterns' risk. It is obvious that cluster #12 and cluster #10 interactive patterns are more concentrated in the lower TTC range than the other clusters. This suggests that the risk level of these two interactive patterns are theoretically higher. Table 2 displays statistics of these two types of interactive patterns. The mean and median of two clusters demonstrate that the risk of cluster #12 is higher than that of cluster #10.

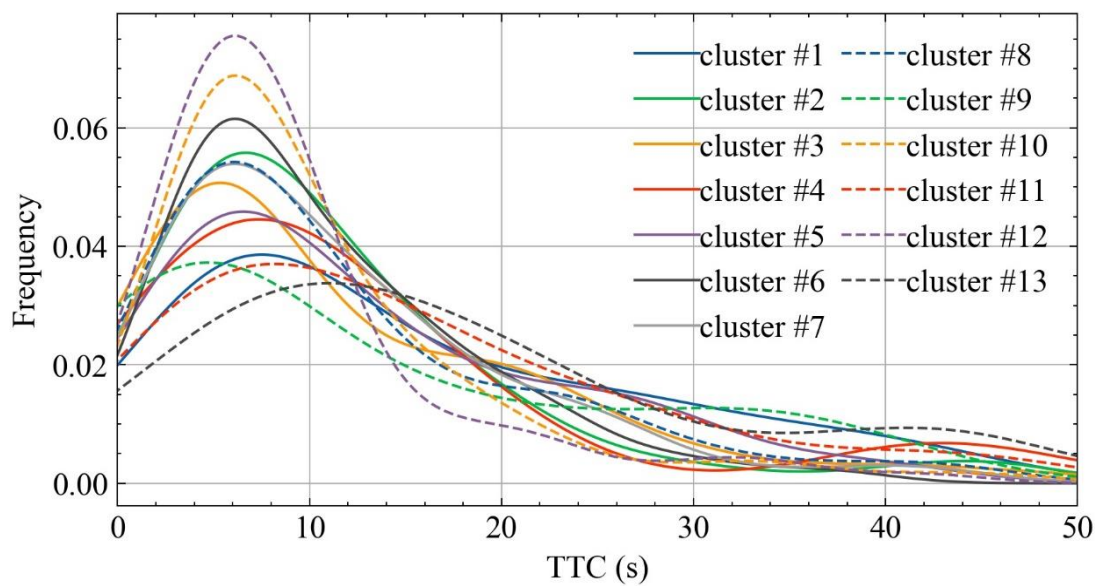

Figure 15 Distribution of TTC observation according to the cluster of interactive primitives

Table 2 TTC statistics of clusters #12 and #10

| Cluster | Frequency | Standard Deviation | Mean | Median |
| --- | --- | --- | --- | --- |
| Cluster #12 | 88 | 7.89 | 9.32 | 7.12 |
| Cluster #10 | 102 | 8.39 | 10.33 | 7.69 |

In order to make in-depth analysis of two clusters of high-risk interactive patterns, Figure 16 shows their dynamic behavior and risk evolution in the interactive process.

Figure 16 (a) represents the later stage of the *ego* vehicle crossing the lane line. Before the *ego* vehicle crosses the lane line, the interactive risk between the *ego* and the *por* vehicles is high. The high risk comes from the close distance and large speed difference between these two vehicles. There is also interactive risk between the *ego* and *ta* vehicles, because their paths cross and the *ego* vehicle's lateral speed is high. During the *ego* vehicle keep away from the lane line, there is no interaction between the *ego* and *por* vehicles, but only between the *ego* and *ta* vehicles. However, the interactive risk between the *ego* and *ta* vehicles is low because of the large spacing.

Figure 16 (b) shows the beginning stage of LC. At first, there are interactive risk between the *ego* and *por* vehicles (the longitudinal speed of *ego* vehicle is higher than that of *por* vehicle). In order to obtain a higher speed advantage, the *ego* vehicle starts to change lanes when the *ta* vehicle is located near the rear of the target lane. Thus, the *ego* vehicle reduces the longitudinal speed to wait for the chance of LC. During this period, the interactive risk between the *por* and *ego* vehicles decreases slightly. At about 1.5s, the *ego* vehicle no longer decelerates, so that the interactive risk between the *ego* and *por* vehicles continue to increase.

Therefore, the extracted driving primitives can also be used to deeply understand the mechanism and evolution of risk in LC interactive patterns.

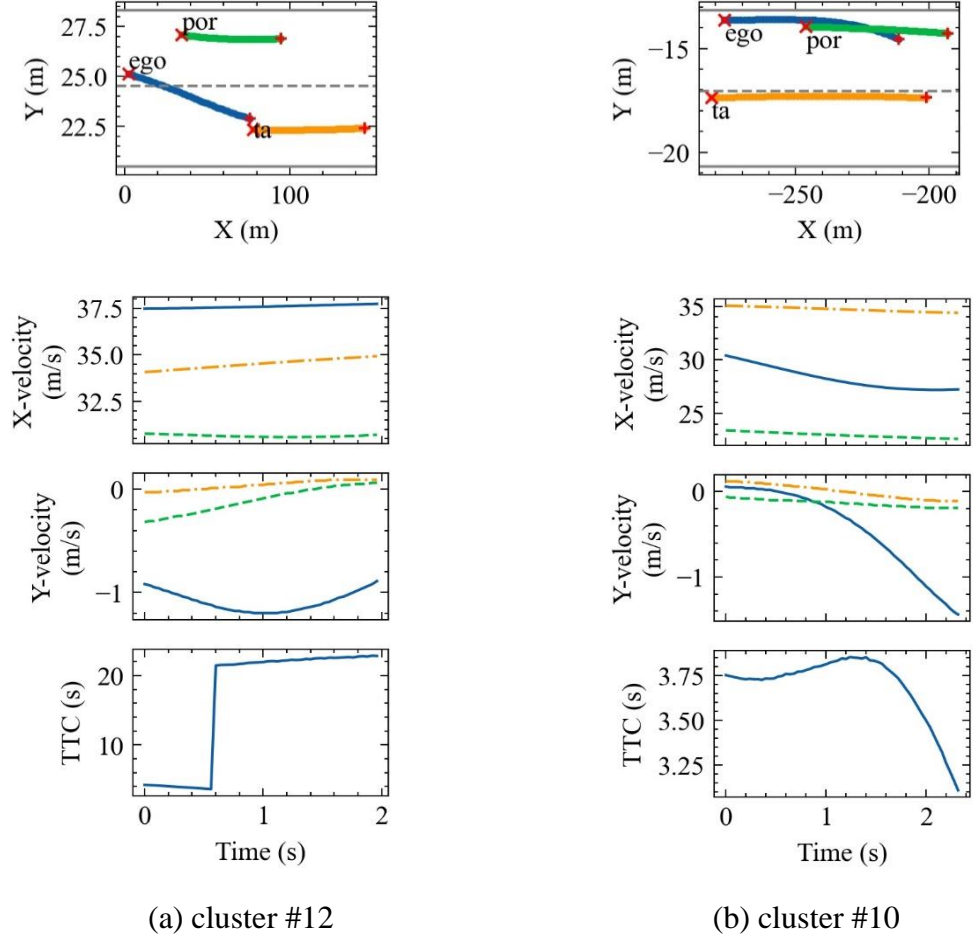

(a) cluster #12    (b) cluster #10

Figure 16 Trajectory, speed and risk evolution graph from two randomly selected clusters.

## 5. Conclusion

This study proposes an unsupervised learning framework that combines primitive-based interactive pattern recognition methods and risk analysis methods. The GMM-HMM is used to decompose LC interactive scenarios into primitives, and TTC is used as an indicator for risk analysis. According to the experimental results of highD naturalistic driving dataset, the following findings are obtained:

(1) The complete LC interactive scenarios can be segmented into interpretable primitives, thereby identifying finite kinds of understandable interactive patterns.

(2) The interactive patterns reproduce the real interactive scenarios and reflect the interaction mechanism and evolution law.

(3) The proposed framework is suitable to analyze the high-risk interactive patterns in LC scenario and explain the risk formation process.

Therefore, the framework proposed in the study can be applied to analyze interactive patterns, understand human behavior, and provide prior knowledge for autonomous vehicle decisions, thereby promoting safe and smooth interaction between autonomous vehicles and human vehicles.

The interactive pattern recognition and risk analysis framework proposed in this paper is flexible in form and accessible to add or delete components. The proposed methodology can be extended to other time-series analysis such as traffic flow data and crash data. In addition, it is also suitable for identifying other scenarios involving multiple traffic participants, such as pedestrian-bicycle-vehicle interaction at intersections and urban roads. This study only explores the interactive patterns between manually driven vehicles. In the future, taking autonomous vehicles into consideration and analyzing the interactive pattern under mixed traffic flow conditions will help autonomous vehicles to make more accurate decisions.

## Acknowledgments

This research was funded by the National Natural Science Foundation of China (Grant No. 71971160) and the Shanghai Science and Technology Committee (Grant No. 19210745700).